# The 2023 Dengue Outbreak in Lombardy, Italy: A One-Health Perspective

Francesca Rovida[1,2,*], Marino Faccini[3,*], Carla Molina Grané[4,*] Irene Cassaniti[1,2], Sabrina Senatore[3], Eva Rossetti[3], Giuditta Scardina[3], Manuela Piazza[5], Giulia Campanini[2], Daniele Lilleri[2], Stefania Paolucci[2], Guglielmo Ferrari[2], Antonio Piralla[2], Francesco Defilippo[6], Davide Lelli[6], Ana Moreno[6], Luigi Vezzosi[7,8], Federica Attanasi[8,], Soresini Marzia[3], Barozzi Manuela[3], Lorenzo Cerutti[9], Stefano Paglia[10], Angelo Regazzetti[5], Maurilia Marcacci[11], Guido Di Donato[11], Marco Farioli[8], Mattia Manica[4,12], Piero Poletti[4,12], Antonio Lavazza[6], Maira Bonini[3], Stefano Merler[4,12,**], Fausto Baldanti[1,2,**], Danilo Cereda[8,**], Lombardy Dengue network

[1.] Department of Clinical, Surgical, Diagnostic and Paediatric Sciences, University of Pavia, Pavia, Italy
[2.] SC Microbiology and Virology, IRCCS Policlinico San Matteo, Pavia, Italy
[3.] Department of Hygiene and Health Prevention, Health Protection Agency, Metropolitan Area of Milano, Milano, Italy
[4.] Center for Health Emergencies, Bruno Kessler Foundation, Trento, Italy
[5.] SC Infectious and Tropical Diseases -ASST Lodi, Lodi, Italy
[6.] Virology Unit, Animal Health and Welfare Department, Istituto Zooprofilattico Sperimentale della Lombardia e dell'Emilia Romagna, Brescia, Italy
[7.] Department of Hygiene and Health Prevention, Health Protection Agency Val Padana, Mantova, Italy
[8.] General Directorate of Welfare, Regione Lombardia, Milano, Italy
[9.] SC Chemical-Clinical Analysis and Microbiology Laboratory, ASST Lodi, Lodi, Italy
[10.] Department of Emergency and Urgency, ASST Lodi, Lodi, Italy
[11.] Istituto Zooprofilattico Sperimentale dell'Abruzzo e Molise, Teramo, Italy
[12.] Epilab-JRU, FEM-FBK Joint Research Unit, Trento, Italy

\* These authors contributed equally to this work and share first authorship.
\*\* These authors contributed equally to this work and share last authorship.

**Correspondence to:** Irene Cassaniti, irene.cassaniti@unipv.it; i.cassaniti@smatteo.pv.it

**Lombardy Dengue network:** Antonella Sarasini, Milena Furione, Dalila Mele, Federica Bergami, Josè Camilla Sammartino, Alessandro Ferrari, Greta Romano, Antonino Maria Guglielmo Pitrolo (SC Microbiology and Virology, IRCCS Policlinico San Matteo, Pavia, Italy); Maya Carrera (Virology Department, Istituto Zooprofilattico Sperimentale della Lombardia ed Emilia Romagna, Brescia, Italy); Rita Brugnoli, Nunzia Laini, Francesca Bonalda, Sara Arfani, Giuditta Zamboni (Department of Hygiene and Health Prevention, Health Protection Agency, Metropolitan Area of Milano, Milano, Italy); Fanny Delfanti, Piergiuseppe Ferrari, Anxhela Dafa (Department of Emergency and Urgency, ASST Lodi, Lodi, Italy); Antonella Negri, Filippa Parisi (SC Chemical-Clinical Analysis and Microbiology Laboratory, ASST Lodi, Lodi, Italy); Marcello Tirani, Michela Viscardi, Gabriele Del Castillo, Federica Morani, Francesco Scovenna, Sheila Sansebastian, Manuel Maffeo (General Directorate of Welfare, Regione Lombardia, Milano, Italy); Mario Chiari (General Directorate of Welfare, Regione Lombardia, Milano, Italy; Department of Hygiene and Health Prevention, Health Protection Agency Brescia, Brescia, Italy); Enrico Tallarita (Social Health Director, ASST Lodi, Lodi, Italy)

# ABSTRACT

**Introduction**. Here we reported the virological, entomological and epidemiological characteristics of the large autochthonous outbreak of dengue (DENV) occurred in a small village of the Lombardy region (Northern Italy) during summer 2023.

**Methods.** After the diagnosis of the first autochthonous case on 18 August 2023, public health measures, including epidemiological investigation and vector control measures, were carried out. A serological screening for DENV antibodies detection was offered to the population. In the case of positive DENV IgM, a second sample was collected to detect DENV RNA and verify seroconversion. Entomological and epidemiological investigations were also performed. A modeling analysis was conducted to estimate the dengue generation time, transmission potential, distance of transmission, and assess diagnostic delays.

**Results.** Overall, 416 subjects participated to the screening program and 20 were identified as DENV-1 cases (15 confirmed and 5 probable). In addition, DENV-1 infection was diagnosed in 24 symptomatic subjects referred to the local Emergency Room Department for suggestive symptoms and 1 case was identified through blood donation screening. The average generation time was estimated to be 18.3 days (95% CI: 13.1-23.5 days). $R_0$ was estimated at 1.31 (95% CI: 0.76-1.98); 90% of transmission occurred within 500m. Entomological investigations performed in 46 pools of mosquitoes revealed the presence of only one positive pool for DENV-1.

**Discussion.** This report highlights the importance of synergic surveillance, including virological, entomological and public health measures to control the spread of arboviral infections.

# 1. INTRODUCTION

Dengue virus (DENV) is an arbovirus belonging to the *Flaviviridae* family, classified into four serotypes (DENV 1-4), transmitted by *Aedes* mosquitoes. In the last 25 years, dengue has become one of the most important mosquito-borne diseases affecting humans, with approximately 40 million cases of dengue fever yearly [1]. Transmission of DENV depends on the presence of the competent mosquitoes *Ae. aegypti* and *Ae. albopictus* [2] that have shown a vast geographical expansion [3-4]. In Europe, *Ae. albopictus* was first reported in Albania (1979) and then in Italy (1990), France (1999), Belgium (2000), Spain (2003) and subsequently in other European countries [3-5], but no *Ae. aegypti* has been found in Italy so far. Since the summer 2010, an increasing number of autochthonous DENV cases and outbreaks have been reported in European countries [6] including France [7-12], Croatia [13-14], Spain [15] and Italy [16-18].

This report analyzes one of the largest dengue outbreaks occurred in Europe (in a village of ca 4,500 inhabitants in the Lodi province, Lombardy Region) so far. The analysis combines results from epidemiological investigations of cases, entomological surveillance, clinical and virological findings and modelling estimates of reporting delays, transmission potential, and transmission distance associated with the observed outbreak.

# 2. METHODS

## 2.1 Public health measures

In Italy, the *National Plan for Prevention, Surveillance and Response to Arboviruses, Plan 2020-2025* (PNA2020-25) [19], published by the Italian Ministry of Health, is the primary document guiding arboviral infection prevention and control. In the Lombardy Region, the Regional Health Authorities have implemented a surveillance program for arboviruses.

According to the PNA2020-25 [19], all probable and confirmed cases of imported or autochthonous DENV had to be immediately reported to the local and regional Public Health Authorities. Upon notification of a probable or confirmed DENV case, the local Public Health Authorities had to perform an immediate epidemiological investigation and take measures to prevent local transmission of the disease, including entomological investigation and activation of vector control measures within 24 hours after notification, during the period of vector circulation. Furthermore, when a confirmed or probable autochthonous DENV outbreak is identified, it is necessary to increase human surveillance through active case finding and improve entomological investigation, vector control measures, and safety measures for biological materials of human origin. Suspected DENV cases were referred to an emergency department or an infectious disease unit of ASST-Lodi for clinical evaluation and laboratory testing. DENV cases were defined as probable or confirmed following the Commission Implementing Decision EU 2018/945 [20]. A case was defined as probable when anti-DENV IgM was detected in a single serum sample. A case was confirmed if any of the following condition were met: i) virus isolation in culture or viral RNA detection in clinical specimens (blood, urine); ii) positive DENV NS1 antigen in serum or plasma; iii) seroconversion or a fourfold increase in DENV IgG antibody titer in two serum samples collected at least two weeks apart; iv) high IgM titre confirmed by neutralization assays. A case was considered autochthonous if no travel history was reported in the two weeks before symptom onset. Collected data include: (1) classification of confirmed or probable case; (2) date of reporting and of symptom onset of cases; (3) geolocation of likely exposure of cases.

**2.2 Vector control measures**

The Agency for Health Protection of the Metropolitan Area of Milan (ATS CMM) requested the implementation of prevention and control measures for confirmed and probable cases of DENV. The objective of these measures was to limit the circulation of *Aedes* spp. mosquitoes and the spread of the virus, as well as inform the citizens and suggest to them personal behaviors and environmental treatments to adopt to prevent the infection. Vector control measures included: i) adulticide and larvicide treatments carried out within a 200 m radius of the house of each case and the areas they frequented within 24 hours of notification; ii) door-to-door removal of larval breeding sites within a 200 m radius of the house of each case and the areas they frequented; iii) periodic adulticide and larvicide treatments in all areas of the village, especially in green spaces, schools, cemeteries, and public drains; iv) monitoring of the presence of *Aedes* spp. mosquitoes to assess the effectiveness of treatments; v) a specific ordinance for private individuals to implement similar adulticide and larvicide treatments on private properties and to adopt personal behaviors to limit mosquito bites (mosquito nets, skin repellents, etc).

The municipality, in addition to the regularly scheduled larvicide and adulticide interventions from June to September, has conducted additional pest control measures in response to a request from the ATS following the reporting of cases.

**2.3 Entomological investigations**

Specific entomological inspections were carried out within a radius of 200m around the homes and places attended by the DENV cases. The BG-Lure cartridge (Biogents) method, recommended by the PNA2020-25, was used to collect adult mosquitoes. The BG-trap attempts to mimic convection currents created by human body heat with a fan and mimics human odors through the BG lure (ammonia, caproic acid, and lactic acid) and an octanol lure [21]. The contrasting black and white markings on the trap also provide visual cues that may be attractive to mosquitoes [22].

Field collections were undertaken from 24 August to 5 November 2023. Samples were collected in neighborhoods that reported cases of DENV, and traps were placed in residents' homes or on the sides of houses within a 200m radius of a DENV case. Overall, 22 sites were investigated, and the mosquitoes were trapped during the night from 2 PM until 9 AM the following day, every four days. Daily rainfall, temperature parameters (minimum, average, and maximum), and average relative humidity were obtained from the nearest weather station, located about 4 km from the municipality where the DENV cases occurred. Data were aggregated by epidemiological weeks to consider cumulative rainfall and weekly average temperature and relative humidity values.

Mosquitoes were frozen and later analysed at Istituto Zooprofilattico Sperimentale della Lombardia e Emilia-Romagna (IZSLER). Only *Ae. albopictus* were collected, sexed, and identified as a species according to morphological taxonomic keys [23-24]. The pools (1 ~ 44 per sample tube) were formed according to sex, data collection, and site. The males were analyzed for transovarial transmission. Sampling tubes containing *Ae. albopictus* were stored at -80°C until testing for viral RNA.

A total of 471 *Ae. albopictus* females were sampled, divided into 42 pools and processed to detect the presence of DENV. The mosquito pools were homogenized, and viral RNA was extracted using the QIAsymphony DNA & RNA Purification System (QIAGEN, Germany) and analyzed using the same pan-Flavivirus heminested RT-PCR targeting the NS5 gene as used for the human samples [25]. The RT-PCR amplification products were sequenced using Sanger sequencing.

**2.4 DENV serological screening**

To actively search for DENV cases in accordance with the PNA2020-25, DENV serological screening was offered to the population of the municipality, which has approximately 4,500 inhabitants. The screening campaign was conducted from 25 August to 15 September, 2023. Afterwards, individuals interested in DENV screening could still access the test by going to the dedicated blood collection centre. After explaining the objectives of the DENV screening, all volunteers who agreed to participate completed a questionnaire to provide information on the presence of DENV-related symptoms and their travel history to DENV-endemic areas since June 2023. Signed informed consents were obtained from all the participants. Blood samples were collected from each participant and analyzed at Microbiology and Virology Department of the Fondazione IRCCS Policlinico San Matteo, Pavia, for DENV IgM and IgG detection. As shown in the flowchart in Figure 1, individuals who tested positive for DENV IgM were considered to have a suspected DENV infection and were subsequently contacted for epidemiological investigation and infectious disease consultation. Additionally, they were asked to undergo further serological and virological testing (in plasma and urine) to confirm the DENV infection by detecting seroconversion or DENV RNA. For individuals who tested positive for DENV IgG and negative for DENV IgM, indicating potential past exposure to DENV, we collected and verified information on symptoms and travel to DENV endemic areas.

**2.5 Laboratory investigations**

Serum samples were tested for DENV IgM and IgG using a chemiluminescent assay (Dengue VirClia IgM monotest and Dengue VirClia IgG monotest, VirCell Microbiologists). Results were expressed as an index and were given as positive when an index was higher than 1.1. Virological investigations were performed in plasma and urine samples of patients with potential DENV infections. DNA was extracted with QIAsymphony® DSP Virus/Pathogen Kit on QIAsymphony® SP automated platform (QIAGEN, Germany). The detection of DENV RNA was performed using a real-time reverse transcriptase-polymerase chain reaction (RT-PCR) targeting a conserved region of DENV [26] and a pan-Flavivirus heminested RT-PCR [25] followed by sequencing of amplicons.

**2.6 Metagenomic NGS Sequencing**

RNA was treated with TURBO DNase (Thermo Fisher Scientific, Waltham, MA) at 37 °C for 20 min and then purified by RNA Clean and Concentrator-5 Kit (Zymo Research). RNA was used for the assessment of sequencing independent single primer amplification protocol (SISPA) with some modifications reported by Lorusso [27]. Libraries were prepared by using Nextera DNA Flex Library Prep (Illumina Inc., San Diego, CA) according to the manufacturer's protocol. Sequencing was performed on the MiSeq (Illumina Inc., San Diego, CA) by MiSeq Reagent v2 (300-cycle). FastaQ obtained were analyzed with the CZ ID metagenomic pipeline [28]. Dengue virus consensus sequences were obtained mapping to the reference with the major coverage breadth and coverage deep obtained by the metagenomic pipeline.

**2.7 Epidemiological methods**

The diagnostic delay was defined as the number of days between the date of symptom onset and the date of diagnosis. We classified each case as occurring before or after the detection of the outbreak on 18 August 2023, according to their date of symptom onset. We provided descriptive statistics (mean, interquartile range); we applied a Poisson regression model (based on log link) to estimate the potential reduction in the diagnostic delay for cases occurring after the outbreak detection.

We used a transmission model based on a previously described Bayesian approach [29-32] to infer likely transmission chains in Lodi province's village involved in the DENV outbreak from geolocated data of identified autochthonous cases. This allowed us to estimate the DENV generation time (i.e., the distribution of the time between successive human infections in a transmission chain) and the transmission distance of the infection (i.e., the distribution of the distance between the residence of infected and their infectors).

Leveraging on the estimated generation time, we quantified the DENV transmission potential in terms of the basic and net reproduction number [29, 33]. The basic reproduction number $R_0$ is defined as the average number of secondary cases generated by a primary infector in a fully susceptible population and under constant epidemiological conditions. The net reproduction number $R_t$ quantifies the transmission at any given time *t*, accounting for the impact of control interventions and seasonal variations in the vector density. Details in the Supplementary Materials.

## 3. RESULTS

### 3.1 Clinical and virological results

The first confirmed case developed initial symptoms on 3 August and was notified on 18 August. Clinical characteristics are reported in a previous article [17]. As of 15 November 2023, a total of 45 cases had been reported, with 40 confirmed cases and five probable cases. Out of these cases, 24 were individuals who presented dengue-like symptoms and were confirmed as DENV cases by laboratory testing after consulting the Emergency Room or Infectious Disease Unit of ASST-Lodi hospitals while 20 cases were detected through DENV serological screening, and one case was identified through DENV NAT testing during a blood donation. All the 24 cases identified through medical consultations tested positive for DENV RNA in plasma (20/24, 83%) and/or urine (12/24, 50%). A subsequent Sanger sequencing analysis performed on plasma and/or urine tested positive with pan-flavivirus heminested RT-PCR revealed the presence of DENV-1 in 22 cases out of 24 (92%). Otherwise, among the cases identified through serological screening program, 9/20 (45%) tested positive for DENV RNA in plasma and/or urine and 6/20 (30%) were confirmed by seroconversion. All nine molecularly positive cases were found to be DENV-1. Five cases were not confirmed through molecular and/or serological approaches and were defined as "probable". Complete genome was performed in 13 DENV-1 strains (five reported by Cassaniti et al. [17] and eight newly sequenced) and one mosquito pool. The Italian strains clustered with Peruvian strains collected from 2021 to 2023, with an average nucleotide identity of 99.2% (range: 98.0–99.7) between DENV-1 strains of genotype V. Moreover, Italian strains had an average nucleotide identity of 99.6% (range: 99.5–99.7) with mosquito pools. The 45 DENV cases had a median age of 62 years (range: 3-95) with a male-to-female ratio of 2:1. Forty-one (91%) were symptomatic, while the remaining four (9%), identified through serological screening, were asymptomatic. The most frequently reported symptoms were fever, asthenia, headache, rash, myalgia, arthralgia, retro-orbital pain and digestive disorders (Table 1). The duration of symptoms ranged from 2 to 28 days, with a median duration of 10 days. Hospitalization was required for 7/41 (15%) cases. Three cases were hospitalized due to persistent symptoms, while four were admitted as a precautionary measure. All cases resided in the same village, except for two who lived in two nearby municipalities and had visited the village during the DENV outbreak period. All cases underwent clinical evaluation. Furthermore, the patients were informed that they were at risk of more severe DENV infection if they travelled to endemic countries in the future and about the opportunity to be vaccinated for DENV before a trip in endemic areas.

**3.2 DENV serological screening**

Overall, 416 subjects participated in the screening, representing about 9% (416/4478) of the population (median age was 58 years; range 2-95). Among them, 35 (9%) individuals tested positive for DENV-specific IgM and/or IgG; in details, 25/35 (71%) tested positive for DENV IgM, 10/35 (29%) tested positive exclusively for DENV IgG. None of the individuals with positive results for DENV IgG reported traveling to dengue-risk areas during the investigated period. Among the 25 individuals with positive results for DENV IgM, two refused further examinations, remaining classified as probable cases, while 23 subsequently underwent additional diagnostic tests. Of these, 15 were confirmed as DENV infections, five were ruled out as non-cases, and three remained probable, resulting in five probable cases overall. Among the 20 cases identified through the community screening, 16 reported having symptoms compatible with DENV before the test, with an average delay between symptom onset and sample collection of 16 days, whereas four cases consistently showed no symptoms.

### 3.3 Entomological investigations and vector control measures

A total of 514 *Ae. albopictus* mosquitoes (471 females and 43 males) were collected. The temperatures, rainfall, and relative humidity were typical of the Mediterranean area: mean week temperatures fluctuated between 29.1°C and 20°C, precipitation was very low and only occurred in the periods 28 August-3 September and 18 September-1 October 2023, the relative humidity fluctuated between a minimum of 59.4% to a maximum of 82.7%. The average temperature, rain, and relative humidity did not apparently affect the number of mosquitoes sampled; instead, mosquito catches were abundant above a minimum temperature of about 16 °C (Figure 2 A-C).

Of the 42 female and 4 male pools tested, only one was positive for DENV. The positive pool was composed of 20 female mosquitoes collected on 2 September 2023 at the working site of the second case of DENV. The pools composed of male mosquitoes were all negative, and the vertical transmission of DENV in *Ae. Albopictus* was not detected in this study. Sequencing of the mosquito pool that resulted positive confirmed the presence of DENV-1.

### 3.4 Epidemiology

The average delay between symptom onset and diagnosis between 1 July and 20 October 2023, was 17.7 days (median: 15 IQR: 5-24 days). The delay decreased from 27.2 days (median: 23.0 IQR: 17-29 days) before the detection of the outbreak on 18 August to 13.4 days (median: 6.5 IQR: 3-21 days) afterward, with an average 50.8% reduction of the diagnostic delay.

The generation time was estimated to be gamma-distributed with a mean 18.3 days (95% CI: 13.1-23.5 days) and a standard deviation of 8.1 days (95% CI: 5.6-9.9 days). $R_0$ was estimated at 1.31 (95% CI: 0.76-1.98). The net reproduction number $R_t$ was found to peak (1.86, 95% CI: 1.26-2.57) on 19 August, just before vector control interventions were implemented (Fig. 3A). The upper 95% CI of $R_t$ decreased below 1 on 10 September 2023, and it remained below the epidemic threshold afterward. The average transmission distance was estimated at 272 m (95% CI: 215-331 m), with about 90% of transmission events occurring within 500m (Fig. 3B).

Similar results were obtained by considering only cases confirmed via PCR and by considering alternative assumptions on the distribution of the dengue generation time (see Supplementary Materials).

## 4. DISCUSSION

This report describes the human, entomological, and epidemiological investigations and public health control measures carried out during an outbreak of DENV-1 in the Lombardy Region from August to November 2023. The outbreak involved 45 autochthonous cases (40 confirmed and 5 probable). All cases were either residents of the same village in the province of Lodi or individuals who had visited the village during the period of DENV circulation. The index case has not been identified despite continued active case finding. It has been speculated that the index case may have been an imported viraemic case returning from an area where DENV is endemic. This hypothesis is further supported by the fact that the likelihood of DENV being introduced by viremic travelers is thought to be significantly higher than the importation of DENV in mosquitoes [34-35].

Upon identification of the first case, the local health authorities promptly conducted an epidemiological investigation and implemented measures to prevent local transmission of the disease. These included entomological surveys and vector control measures, which were repeated after each case was detected. The timely implementation of containment measures may have helped limit the spread of the virus to neighboring areas by containing the outbreak within the same village.

Voluntary serological screening for DENV offered to the population of the village aimed to increase human surveillance by actively searching for cases, following the diagnosis of the first cases and led to the identification of 20 DENV cases. Otherwise, 24 individuals presented dengue-like symptoms were confirmed as DENV cases by laboratory testing after medical consultation and one case was identified through DENV NAT testing during a blood donation.

About clinical signs, 91% of identified cases of DENV were symptomatic and presented non-severe symptoms, such as fever, rash, headache, asthenia, arthralgia, and myalgia. Of these cases, 15% required hospitalization, and no deaths or severe disease cases were reported. Despite the proportion of symptomatic cases being much higher than the global estimate reported in the literature [36], this can be partially attributed to the fact that individuals who had recently experienced symptoms compatible with dengue were more likely to adhere to serological screening compared to the rest of the population. Moreover, according to a recent systematic review, although the average proportion of inapparent DENV infections is estimated to be globally 80% (95% CI: 72 -88), it may vary greatly across geographic areas, ranging from 19% (95% CI: 17–21) in the Eastern Mediterranean region to 93% (95% CI: 89-98) in the Southeast Asia region [37]. Given that DENV infections are often asymptomatic [38] or present with non-specific symptoms and that only 9% of the population residing in the affected village participated to the screening, it is likely that the overall number of DENV case occurred in the area has been underestimated. Molecular investigations confirmed that the outbreak was sustained by DENV-1. This serotype was also responsible for the autochthonous DENV outbreak reported in the Veneto region in 2020 [39] and in the Lazio region in 2023 [18].

In the last decades, an increasing number of autochthonous DENV cases have occurred in Europe where competent *Ae albopictus* is present [7,9,16,18]. In Italy, *Ae albopictus* was reported for the first time in the 1990s [40], and it is today present throughout Italy. In the village affected by DENV-1 outbreak, entomological surveys revealed a conspicuous number of *Ae albopictus*. Among the *Ae albopictus* mosquito polls collected during the entomological surveillance period, one tested positive for DENV-1 and sequences from mosquito pools and humans were almost identical. This data provides further evidence on the potential role of *Ae. albopictus* in sustaining future DENV outbreaks in Europe and emphasizes the risk of future events.

By implementing a modelling Bayesian approach to reconstruct likely transmission chains of DENV between identified cases, we estimated a mean generation time of 18.3 days, in agreement with previous estimates (15 to 19 days) from dengue epidemics occurred outside Europe and sustained mainly by *Aedes aegypti* [29,41]. As expected for temperate climate areas, the basic reproduction number $R_0$ of 1.31 was found to be lower than what estimated in tropical and subtropical areas [3,42]. The relatively low value of $R_0$, combined with a rather long generation time (~3 weeks) and a relatively short duration of the mosquitoes' season in the region [3], likely hindered the spatial spread of DENV-1 in the region while suggesting a low risk of experiencing large epidemics. Our analysis suggests that the identification of the outbreak and consequent vector control interventions quickly reduce the transmissibility of DENV, leading to a reduction of about 50% of the diagnostic delays and bringing $R_t$ below the threshold of 1 in about two weeks. We found that most of the transmission occurred within human walking-distance, with about 90% of transmission events occurring within 500 meters. This result is in line with estimates associated with dengue transmission in Brazil [29] and with the chikungunya outbreak occurred in 2017 in central Italy [30].

The results presented in this study should be interpreted, considering the following limitations. The data collected at the early phase of the outbreak may be incomplete because of the situation of emergency. The generation time was estimated considering only symptomatic infections among the village's residents, therefore assuming that the proportion of asymptomatic cases is low. This assumption is in line with the low number of asymptomatic cases (4 cases) identified with serological screening but contrasts with evidence from endemic regions [43]. More efforts are therefore required to investigate the symptomatic ratio of DENV in non-endemic regions.

It is important to note that the majority of DENV cases in Italy and Europe are imported from endemic areas, particularly during the summer months. Therefore, it is crucial to implement effective measures to prevent both the introduction and the spread and importation of the viral infection during the period of high density and activity of *Aedes* mosquitoes. In particular, in order to minimize the risk of importing the dengue virus into non-endemic areas, it is essential to inform the community about the risk of dengue in travelers and promote seeking specific risk information related to season and travel destinations, alongside necessary precautions, trough pre-travel counseling in specialized travel medicine centers. Moreover, with the development of the new tetravalent live attenuated dengue vaccine QDENGA® (TAK-003), approved for use and commercialization in Italy in February 2023, there is now a new tool available for prevention among travelers to dengue-risk areas.

To reduce the risk of viral infection spreading during *Aedes* high activity months, it is crucial to train healthcare professionals in the early detection of cases and to raise awareness among the general population and other stakeholders (e.g., municipal and local authorities) about the preventive and control measures that have to be taken both at the individual and community level.

## 5. CONCLUSIONS

The number of autochthonous DENV outbreaks in Italy has increased in the last years, with multiple unrelated events and three separate clusters observed over the last summer season [17,18,44]. This highlights the importance of continued integrated surveillance for early detection of imported virus infections and preventing potential outbreaks. Further monitoring should be implemented in order to early verify the possible circulation of infected mosquitoes. In addition, it will be crucial to conduct virological surveillance of *Aedes* mosquitoes to enable early identification of circulating arboviruses and prompt definition of local and regional public health measures, including population awareness that can control or prevent future events. In particular, considering that a large part of larval breeding sites is estimated to be located in domestic or peri-domestic areas (e.g., private gardens, artificial fountains and pools) [45], the engagement of the community for their removal through regular cleaning and maintenance of domestic water storage containers and proper disposal of solid waste is a fundamental pillar of effective mosquito control and prevention efforts. This study, set against the backdrop of a non-typical dengue environment, not only sheds light on the adaptability of pathogens and their vectors but also underscores the importance of preparedness and adaptability in public health strategies in a One-health approach, providing key quantitative metrics describing dengue transmission in Europe. The study's findings could redefine future responses to vector-borne diseases worldwide, particularly in regions previously considered at low risk.

## FIGURE

**Figure 1.** Decisional algorithm for serological and virological screening offered to population. The flow chart includes also epidemiological investigations for definition of autochthonous or imported case and linking to the outbreak.

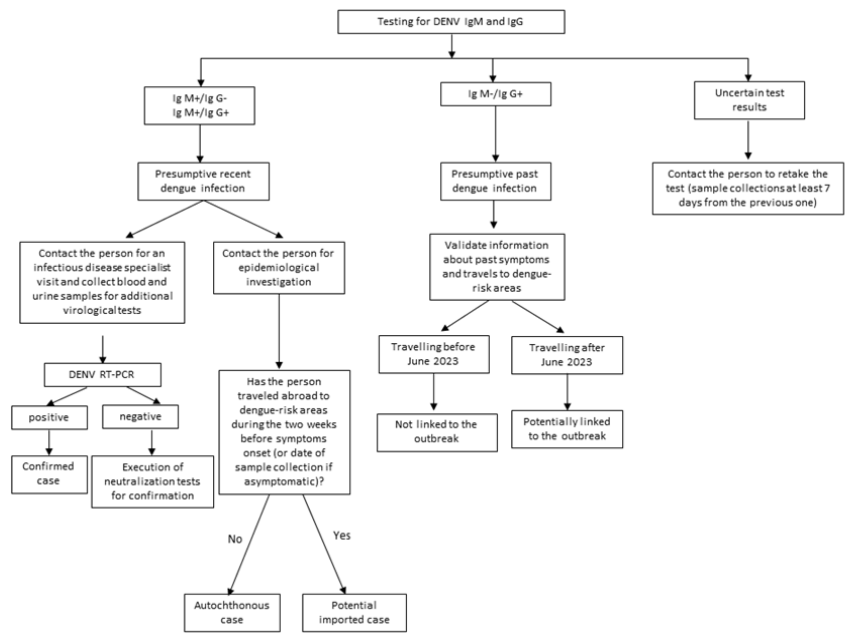

**Figure 2**. Collection of Ae. Albopictus mosquitoes (number of collected samples) divided per week, on the basis of temperature (mean and minimum; A), percentage of rainfall (B) and relative humidity (C).

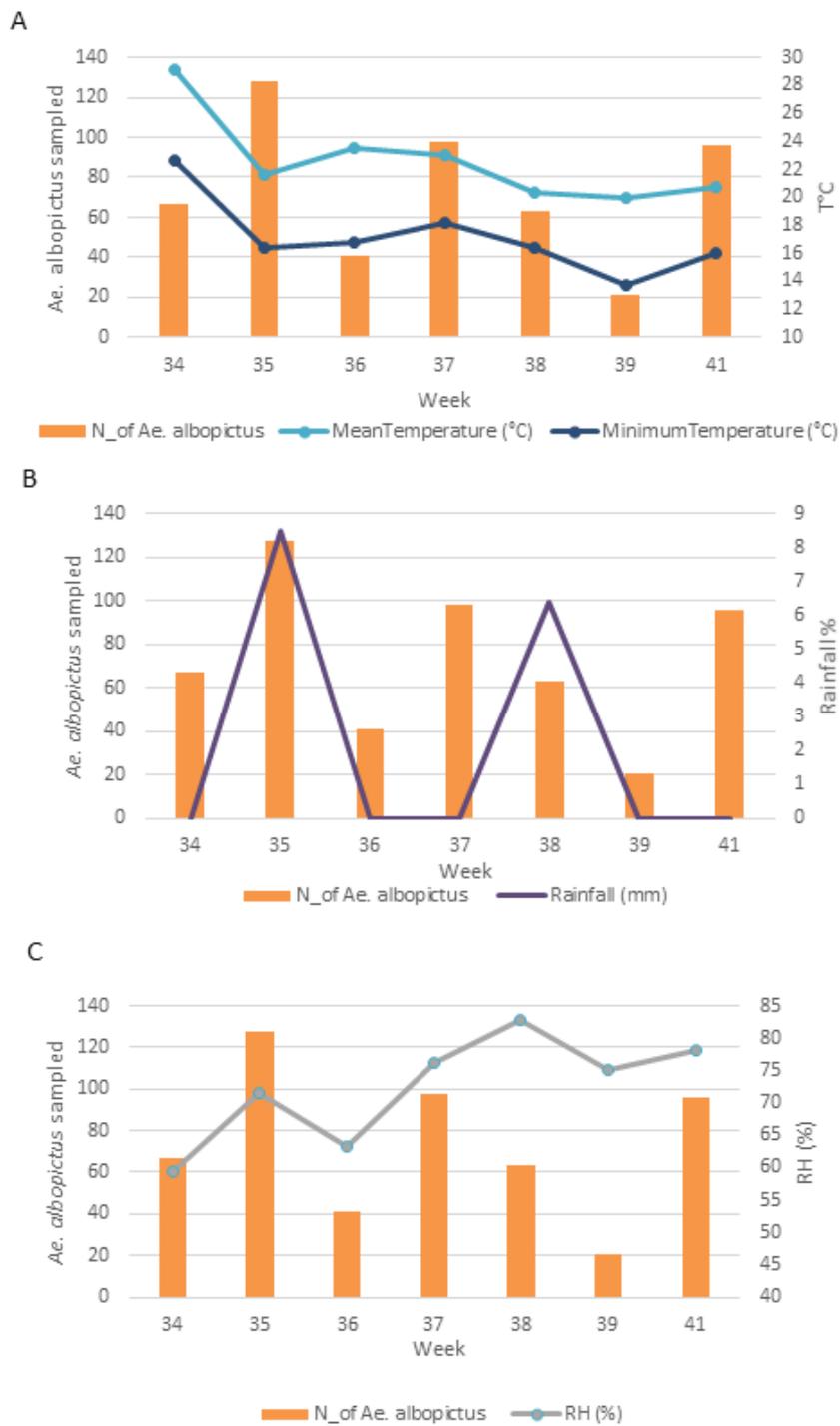

**Figure 3.** A Net reproduction number Rt (the red solid line represents the mean estimate; the shaded area indicates 95% credible intervals; Rt averaged over a moving window of two weeks), scale on the left, from the curve of cases by date of symptom onset (light blue bars), scale on the right. B Estimated cumulative percentage of transmission events as a function of the distance between the residence of infected and of their infectors. Bars represent mean estimates; vertical lines represent 95% credible intervals.

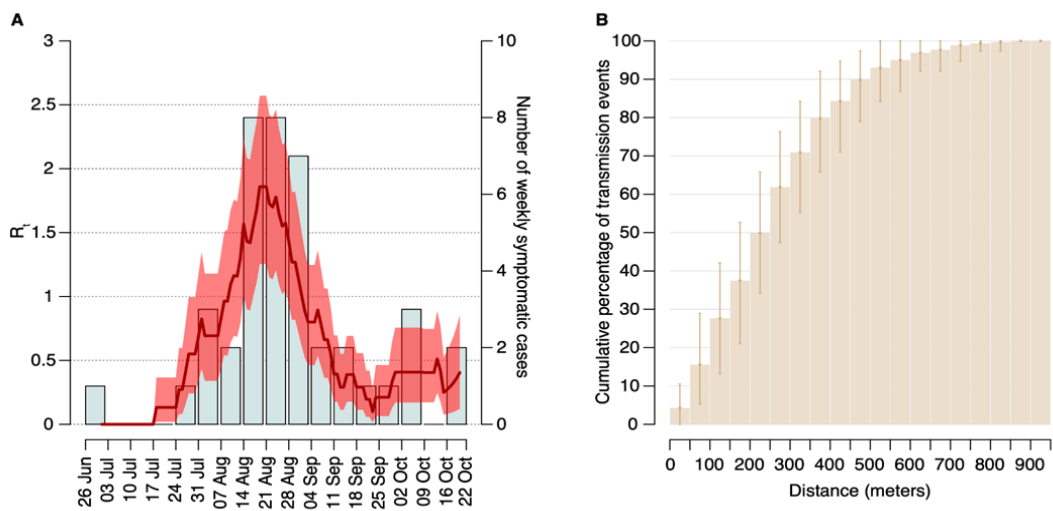

# TABLE

**Table 1**. Demografic and clinical data of 45 autochthonous dengue cases, Lombardy Region, Italy, August-November 2023

|  | DENV cases |
|---|---|
| Demographic characteristics (n=45) |  |
| Female | 15 (33%) |
| Male | 30 (67%) |
| Median age, years | 62 |
| Hospitalisation: <br> Yes <br> No <br> Symptomatic cases <br> Asymptomatic cases | 7 (15%) <br> 38 (85%) <br> 41 (91%) <br> 4 (9%) |
| Symptoms (n=41) |  |
| Fever | 40 (98%) |
| Asthenia | 35 (85%) |
| Headache | 21 (51%) |
| Rash | 21 (51%) |
| Myalgia | 20 (49%) |
| Retro-orbital pain | 19 (46%) |
| Arthralgia | 18 (44%) |
| Digestive disorders | 12 (29%) |

**Funding**

This research was partially supported by EU funding within the NextGeneration EU-MUR PNRR Extended Partnership initiative on Emerging Infectious Diseases (Project no. PE00000007, INF-ACT), Centre for Disease Prevention and Control (CCM; SURVEID Project, program 2022) and Ricerca Corrente (grant no. 80206) of the Italian Ministry of Health.

**Conflict of interest**

The authors have declared no conflicts of interest.

**Acknowledgments**

We thank Daniela Sartori for manuscript editing.

**REFERENCES**

[1] Jelinek T. Trends in the epidemiology of dengue fever and their relevance for importation to Europe. Euro Surveill. 2009;14(25):19250. PMID: 19555595.

[2] Kraemer MUG, Reiner RC Jr, Brady OJ, Messina JP, Gilbert M, Pigott DM et al. Past and future spread of the arbovirus vectors Aedes aegypti and Aedes albopictus. Nat Microbiol. 2019; 4(5):854-863. doi: 10.1038/s41564-019-0376-y. PMID: 30833735.

[3] Zardini A, Menegale F, Gobbi A, Manica M, Guzzetta G, d'Andrea V, Marziano V, Trentini F, Montarsi F, Caputo B, Solimini A, Marques-Toledo C, Wilke ABB, Rosà R, Marini G, Arnoldi D, Pastore Y Piontti A, Pugliese A, Capelli G, Della Torre A, Teixeira MM, Beier JC, Rizzoli A, Vespignani A, Ajelli M, Merler S, Poletti P. Estimating the potential risk of transmission of arboviruses in the Americas and Europe: a modelling study. Lancet Planet Health. 2024 Jan;8(1):e30-e40. doi: 10.1016/S2542-5196(23)00252-8. PMID: 38199719.

[4] Tomasello D, Schlagenhauf P. Chikungunya and dengue autochthonous cases in Europe, 2007-2012. Travel Med Infect Dis. 2013 Sep-Oct;11(5):274-84. doi: 10.1016/j.tmaid.2013.07.006. Epub 2013 Aug 17. PMID: 23962447.

[5] Bellini R, Michaelakis A, Petrić D, Schaffner F, Alten B, Angelini P, Aranda C, Becker N, Carrieri M, Di Luca M, Fălcuţă E, Flacio E, Klobučar A, Lagneau C, Merdić E, Mikov O, Pajovic I, Papachristos D, Sousa CA, Stroo A, Toma L, Vasquez MI, Velo E, Venturelli C, Zgomba M. Practical management plan for invasive mosquito species in Europe: I. Asian tiger mosquito (Aedes albopictus). Travel Med Infect Dis. 2020 May-Jun;35:101691. doi: 10.1016/j.tmaid.2020.101691. Epub 2020 Apr 22. PMID: 32334085.

[6] European Centre for Disease Prevention and Control. Autochthonous vectorial transmission of dengue virus in mainland EU/EEA, 2010–present [cited 2023 Oct 31]. https://www.ecdc.europa.eu/en/all-topics-z/dengue/surveillance-and-disease-data/autochthonous-transmissiondengue-virus-eueea.

[7] La Ruche G, Souarès Y, Armengaud A, Peloux-Petiot F, Delaunay P, Desprès P, Lenglet A, Jourdain F, Leparc-Goffart I, Charlet F, Ollier L, Mantey K, Mollet T, Fournier JP, Torrents R, Leitmeyer K, Hilairet P, Zeller H, Van Bortel W, Dejour-Salamanca D, Grandadam M, Gastellu-Etchegorry M. First two autochthonous dengue virus infections in metropolitan France, September 2010. Euro Surveill. 2010;15(39):19676. PMID: 20929659.

[8] Marchand E, Prat C, Jeannin C, Lafont E, Bergmann T, Flusin O, Rizzi J, Roux N, Busso V, Deniau J, Noel H, Vaillant V, Leparc-Goffart I, Six C, Paty MC. Autochthonous case of dengue in France, October 2013. Euro Surveill. 2013 Dec 12;18(50):20661. doi: 10.2807/1560-7917.es2013.18.50.20661. PMID: 24342514.

[9] Succo T, Leparc-Goffart I, Ferré JB, Roiz D, Broche B, Maquart M, Noel H, Catelinois O, Entezam F, Caire D, Jourdain F, Esteve-Moussion I, Cochet A, Paupy C, Rousseau C, Paty MC, Golliot F. Autochthonous dengue outbreak in Nîmes, South of France, July to September 2015. Euro Surveill. 2016;21(21). doi: 10.2807/1560-7917.ES.2016.21.21.30240. PMID: 27254729.

[10] Succo T, Noël H, Nikolay B, Maquart M, Cochet A, Leparc-Goffart I, Catelinois O, Salje H, Pelat C, de Crouy-Chanel P, de Valk H, Cauchemez S, Rousseau C. Dengue serosurvey after a 2-month long outbreak in Nîmes, France, 2015: was there more than met the eye? Euro Surveill. 2018;23(23):1700482. doi: 10.2807/1560-7917.ES.2018.23.23.1700482. PMID: 29897042; PMCID: PMC6152166.

[11] Cochet A, Calba C, Jourdain F, Grard G, Durand GA, Guinard A; Investigation team; Noël H, Paty MC, Franke F. Autochthonous dengue in mainland France, 2022: geographical extension and incidence increase. Euro Surveill. 2022 Nov;27(44):2200818. doi: 10.2807/1560-7917.ES.2022.27.44.2200818. PMID: 36330819; PMCID: PMC9635021.

[12] Zatta M, Brichler S, Vindrios W, Melica G, Gallien S. Autochthonous Dengue Outbreak, Paris Region, France, September-October 2023. Emerg Infect Dis. 2023 Dec;29(12):2538-2540. doi: 10.3201/eid2912.231472. Epub 2023 Nov 15. PMID: 37967048.

[13] Schmidt-Chanasit J, Haditsch M, Schoneberg I, Gunther S, Stark K, Frank C. Dengue virus infection in a traveller returning from Croatia to Germany. Euro Surveill. 2010;15(40):19677. doi: 10.2807/ese.15.40.19677-en. PMID: 20946759.

[14] Gjenero-Margan I, Aleraj B, Krajcar D, Lesnikar V, Klobučar A, Pem-Novosel I, Kurečić-Filipović S, Komparak S, Martić R, Duričić S, Betica-Radić L, Okmadžić J, Vilibić-Čavlek T, Babić-Erceg A, Turković B, Avsić-Županc T, Radić I, Ljubić M, Sarac K, Benić N, Mlinarić-Galinović G. Autochthonous dengue fever in Croatia, August-September 2010. Euro Surveill. 2011;16(9):19805. PMID: 21392489.

[15] Monge S, García-Ortúzar V, López Hernández B, Lopaz Pérez MÁ, Delacour-Estrella S, Sánchez-Seco MP, Fernández Martinez B, García San Miguel L, García-Fulgueiras A, Sierra Moros MJ; Dengue Outbreak Investigation Team. Characterization of the first autochthonous dengue outbreak in Spain (August-September 2018). Acta Trop. 2020;205:105402. doi: 10.1016/j.actatropica.2020.105402. Epub 2020 Feb 20. PMID: 32088276.

[16] Lazzarini L, Barzon L, Foglia F, Manfrin V, Pacenti M, Pavan G, Rassu M, Capelli G, Montarsi F, Martini S, Zanella F, Padovan MT, Russo F, Gobbi F. First autochthonous dengue outbreak in Italy, August 2020. Euro Surveill. 2020;25(36):2001606. doi: 10.2807/1560-7917.ES.2020.25.36.2001606. PMID: 32914745; PMCID: PMC7502902.

[17] Cassaniti I, Ferrari G, Senatore S, Rossetti E, Defilippo F, Maffeo M, Vezzosi L, Campanini G, Sarasini A, Paolucci S, Piralla A, Lelli D, Moreno A, Bonini M, Tirani M, Cerutti L, Paglia S, Regazzetti A, Farioli M, Lavazza A, Faccini M, Rovida F, Cereda D, Baldanti F; Lombardy Dengue network; Lombardy Dengue Network. Preliminary results on an autochthonous dengue outbreak in Lombardy Region, Italy, August 2023. Euro Surveill. 2023;28(37). doi: 10.2807/1560-7917.ES.2023.28.37.2300471. PMID: 37707980.

[18] De Carli G, Carletti F, Spaziante M, Gruber CEM, Rueca M, Spezia PG, Vantaggio V, Barca A, De Liberato C, Romiti F, Scicluna MT, Vaglio S, Feccia M, Di Rosa E, Gianzi FP, Giambi C, Scognamiglio P, Nicastri E, Girardi E, Maggi F, Vairo F; Lazio Dengue Outbreak Group; Lazio

dengue Outbreak Group. Outbreaks of autochthonous Dengue in Lazio region, Italy, August to September 2023: preliminary investigation. Euro Surveill. 2023;28(44):2300552. doi: 10.2807/1560-7917.ES.2023.28.44.2300552. PMID: 37917030; PMCID: PMC10623645.

[19] Italian Ministry of Health. National Plan for Prevention, Surveillance and Response to Arboviruses (2020-2025); Ministry of Health, Rome, Italy, 2019.https://www.salute.gov.it/imgs/C_17_pubblicazioni_2947_allegato.pdf

[20] European Commission (EC). Commission Implementing Decision (EU) 2018/945 of 22 June 2018 on the communicable diseases and related special health issues to be covered by epidemiological surveillance as well as relevant case definitions. Brussels: EC; 2018. Available from: https://eur-lex.europa.eu/legal-content/EN/TXT/PDF/?uri=CELEX:32018D0945&from=EN#page=18

[21] Meeraus WH, Armistead JS, Arias JR. Field comparison of novel and gold standard traps for collecting Aedes albopictus in Northern Virginia. J Am Mosq Control Assoc. 2008 Jun;24(2):244-8. doi: 10.2987/5676.1. PMID: 18666532.Huhtamo E, Hasu E, Uzcátegui NY, Erra E, Nikkari S, Kantele A, Vapalahti O, Piiparinen H. Early diagnosis of dengue in travelers: comparison of a novel real-time RT-PCR, NS1 antigen detection and serology. J Clin Virol. 2010 ;47(1):49-53. doi: 10.1016/j.jcv.2009.11.001. Epub 2009 Dec 5. PMID: 19963435.

[22] Farajollahi A, Kesavaraju B, Price DC, Williams GM, Healy SP, Gaugler R, Nelder MP. Field efficacy of BG-Sentinel and industry-standard traps for Aedes albopictus (Diptera: Culicidae) and West Nile virus surveillance. J Med Entomol. 2009 Jul;46(4):919-25. doi: 10.1603/033.046.0426. PMID: 19645298.

[23] Severini F, Toma L, Di Luca M and Romi R. Italian mosquitoes: general information and identification of adults (diptera, culicidae) / le zanzare italiane: generalità e identificazione degli adulti (diptera, culicidae). *Fragmenta entomologica.* 2009; 41(2):213–372. doi: 10.13133/2284-4880/92.

[24] Becker N, Petric D, Zgomba M, Boase C, Madon M, Dahl C, Kaiser A. Mosquitoes and their control. Berlin, Heidelberg: Springer; 2010. https://link.springer.com/book/10.1007/978-3-540-92874-4.

[25] Scaramozzino N, Crance JM, Jouan A, DeBriel DA, Stoll F, Garin D. Comparison of flavivirus universal primer pairs and development of a rapid, highly sensitive heminested reverse transcription-PCR assay for detection of flaviviruses targeted to a conserved region of the NS5 gene sequences. J Clin Microbiol. 2001;39(5):1922-7. https://doi.org/10.1128/ JCM.39.5.1922-1927.2001 PMID: 11326014.

[26] Huhtamo E, Hasu E, Uzcátegui NY, Erra E, Nikkari S, Kantele A, Vapalahti O, Piiparinen H. Early diagnosis of dengue in travelers: comparison of a novel real-time RT-PCR, NS1 antigen detection and serology. J Clin Virol. 2010 ;47(1):49-53. doi: 10.1016/j.jcv.2009.11.001. Epub 2009 Dec 5. PMID: 19963435.

[27] Lorusso A, Calistri P, Mercante MT, Monaco F, Portanti O, Marcacci M, Cammà C, Rinaldi A, Mangone I, Di Pasquale A, Iommarini M, Mattucci M, Fazii P, Tarquini P, Mariani R, Grimaldi A, Morelli D, Migliorati G, Savini G, Borrello S, D'Alterio N. A "One-Health" approach for diagnosis and molecular characterization of SARS-CoV-2 in Italy. One Health. 2020 Apr 19;10:100135. doi: 10.1016/j.onehlt.2020.100135.

[28] Kalantar KL, Carvalho T, de Bourcy CFA, Dimitrov B, Dingle G, Egger R, Han J, Holmes OB, Juan YF, King R, Kislyuk A, Lin MF, Mariano M, Morse T, Reynoso LV, Cruz DR, Sheu J, Tang J, Wang J, Zhang MA, Zhong E, Ahyong V, Lay S, Chea S, Bohl JA, Manning JE, Tato CM, DeRisi JL. IDseq-An open source cloud-based pipeline and analysis service for metagenomic pathogen detection and monitoring. Gigascience. 2020 Oct 15;9(10): giaa111. doi:10.1093/gigascience/giaa111. PMID: 33057676; PMCID: PMC7566497.

[29] Guzzetta G, Marques-Toledo CA, Rosà R, Teixeira M, Merler S. Quantifying the spatial spread of dengue in a non-endemic Brazilian metropolis via transmission chain reconstruction. Nat

Commun. 2018 Jul 19;9(1):2837. doi: 10.1038/s41467-018-05230-4. PMID: 30026544; PMCID: PMC6053439.

[30] Guzzetta G, Vairo F, Mammone A, Lanini S, Poletti P, Manica M, Rosa R, Caputo B, Solimini A, Torre AD, Scognamiglio P, Zumla A, Ippolito G, Merler S. Spatial modes for transmission of chikungunya virus during a large chikungunya outbreak in Italy: a modeling analysis. BMC Med. 2020 Aug 7;18(1):226. doi: 10.1186/s12916-020-01674-y. PMID: 32762750; PMCID: PMC7412829.

[31] Jombart T, Cori A, Didelot X, Cauchemez S, Fraser C, Ferguson N. Bayesian reconstruction of disease outbreaks by combining epidemiologic and genomic data. PLoS Comput Biol. 2014 Jan;10(1):e1003457. doi: 10.1371/journal.pcbi.1003457. Epub 2014 Jan 23. PMID: 24465202; PMCID: PMC3900386.

[32] Lau MS, Dalziel BD, Funk S, McClelland A, Tiffany A, Riley S, Metcalf CJ, Grenfell BT. Spatial and temporal dynamics of superspreading events in the 2014-2015 West Africa Ebola epidemic. Proc Natl Acad Sci U S A. 2017 Feb 28;114(9):2337-2342. doi: 10.1073/pnas.1614595114. Epub 2017 Feb 13. PMID: 28193880; PMCID: PMC5338479.

[33] Cori A, Ferguson NM, Fraser C, Cauchemez S. A new framework and software to estimate time-varying reproduction numbers during epidemics. Am J Epidemiol. 2013 Nov 1;178(9):1505-12. doi: 10.1093/aje/kwt133. Epub 2013 Sep 15. PMID: 24043437; PMCID: PMC3816335.

[34] Wilder-Smith A, Gubler DJ. Geographic expansion of dengue: the impact of international travel. Med Clin North Am. 2008 Nov;92(6):1377-90, x. doi: 10.1016/j.mcna.2008.07.002. PMID: 19061757.

[35] Wilder-Smith A, Quam M, Sessions O, Rocklov J, Liu-Helmersson J, Franco L, Khan K. The 2012 dengue outbreak in Madeira: exploring the origins. Euro Surveill. 2014 Feb 27;19(8):20718. doi: 10.2807/1560-7917.es2014.19.8.20718. Erratum in: Euro Surveill. 2014;19(9):20725. PMID: 24602277.

[36] Bhatt S, Gething PW, Brady OJ, Messina JP, Farlow AW, Moyes CL, Drake JM, Brownstein JS, Hoen AG, Sankoh O, Myers MF, George DB, Jaenisch T, Wint GR, Simmons CP, Scott TW, Farrar JJ, Hay SI. The global distribution and burden of dengue. Nature. 2013 Apr 25;496(7446):504-7. doi: 10.1038/nature12060. Epub 2013 Apr 7. PMID: 23563266; PMCID: PMC3651993.

[37] Li Z, Wang J, Cheng X, Hu H, Guo C, Huang J, Chen Z, Lu J. The worldwide seroprevalence of DENV, CHIKV and ZIKV infection: A systematic review and meta-analysis. PLoS Negl Trop Dis. 2021 Apr 28;15(4):e0009337. doi: 10.1371/journal.pntd.0009337. PMID: 33909610; PMCID: PMC8109817.

[38] Sanchez-Vegas C, Hamer DH, Chen LH, Wilson ME, Benoit C, Hunsperger E, Macleod WB, Jentes ES, Ooi WW, Karchmer AW, Kogelman L, Yanni E, Marano N, Barnett ED. Prevalence of dengue virus infection in US travelers who have lived in or traveled to dengue-endemic countries. J Travel Med. 2013 Nov-Dec;20(6):352-60. doi: 10.1111/jtm.12057. Epub 2013 Aug 16. PMID: 24165381.

[39] Barzon L, Gobbi F, Capelli G, Montarsi F, Martini S, Riccetti S, Sinigaglia A, Pacenti M, Pavan G, Rassu M, Padovan MT, Manfrin V, Zanella F, Russo F, Foglia F, Lazzarini L. Autochthonous dengue outbreak in Italy 2020: clinical, virological and entomological findings. J Travel Med. 2021 Dec 29;28(8):taab130. doi: 10.1093/jtm/taab130. PMID: 34409443; PMCID: PMC8499737.

[40] Romi R. History and updating on the spread of Aedes albopictus in Italy. Parassitologia. 1995;37(2-3):99-103. PMID:8778671.

[41] Sanches RP, Massad E. A comparative analysis of three different methods for the estimation of the basic reproduction number of dengue. Infect Dis Model. 2016 Aug 31;1(1):88-100. doi: 10.1016/j.idm.2016.08.002. PMID: 29928723; PMCID: PMC5963322.

[42] Liu Y, Lillepold K, Semenza JC, Tozan Y, Quam MBM, Rocklöv J. Reviewing estimates of the basic reproduction number for dengue, Zika and chikungunya across global climate zones. Environ Res. 2020 Mar;182:109114. doi: 10.1016/j.envres.2020.109114. Epub 2020 Jan 3. PMID: 31927301.

[43] Montoya M, Gresh L, Mercado JC, Williams KL, Vargas MJ, Gutierrez G, Kuan G, Gordon A, Balmaseda A, Harris E. Symptomatic versus inapparent outcome in repeat dengue virus infections is influenced by the time interval between infections and study year. PLoS Negl Trop Dis. 2013 Aug 8;7(8):e2357. doi: 10.1371/journal.pntd.0002357. PMID: 23951377; PMCID: PMC3738476.

[44] Vita S, Bordi L, Sberna G, Caputi P, Lapa D, Corpolongo A, Mija C, D'Abramo A, Maggi F, Vairo F, Specchiarello E, Girardi E, Lalle E, Nicastri E. Autochthonous Dengue Fever in 2 Patients, Rome, Italy. Emerg Infect Dis. 2024 Jan;30(1):183-184. doi: 10.3201/eid3001.231508. Epub 2023 Nov 15. PMID: 37967518; PMCID: PMC10756386.

[45] Ali EOM, Babalghith AO, Bahathig AOS, Toulah FHS, Bafaraj TG, Al-Mahmoudi SMY, Alhazmi AMF, Abdel-Latif ME. Prevalence of Larval Breeding Sites and Seasonal Variations of *Aedes aegypti* Mosquitoes (Diptera: Culicidae) in Makkah Al-Mokarramah, Saudi Arabia. Int J Environ Res Public Health. 2021 Jul 9;18(14):7368. doi: 10.3390/ijerph18147368. PMID: 34299819; PMCID: PMC8305884

**Supplementary Materials**

**Estimation of the generation time and the transmission distance**

To estimate the generation time and to assess the geographical distance of the transmission events, we developed a spatio-temporal transmission model following the same approach presented in previous studies [1–3]. The model considers the force of infection over time exerted on each case $j$, $\lambda_j(t)$, which accounts for the contribution of the cases infected before $j$, $i \in I_j$, in the same area. This force of infection can be defined as

$$\lambda_j(t) = \sum_{i \in I_j} \lambda_{j,i}(t) = \sum_{i \in I_j} \beta \; e^{-\eta \, d_{j,i}} \, \varphi(t - T_i \, ; \, \mu, \, \sigma^2),$$

where $\beta$ is a scaling parameter shaping the transmissibility of dengue, $e^{-\eta \, d_{j,i}}$ is the distance kernel accounting for the probability that a transmission event occurs at a distance $d_{j,i}$ between the residences of individuals $i$ and $j$, $1/\eta$ is the kernel's characteristic distance, $T_i$ is the time of infection of individual $i$, and $\varphi(t \, ; \, \mu, \, \sigma^2)$ refers to the generation time distribution, which we assumed to be gamma-distributed with parameters mean $\mu$ and variance $\sigma^2$. The generation time accounts for the length of the incubation period in both mosquitoes and humans, for the duration of the infectiousness in humans, and for the life expectancy of an adult mosquito. At each time interval $[t, t + \Delta t]$, a susceptible individual $j$ gets infected with probability $p(t) = 1 - e^{-\lambda_j(t)}$.

Model parameters $\beta, \eta, \mu,$ and $\sigma^2$ are unknown, and the infection times $T_i$ of the identified cases are unobserved events. Both sets of parameters are estimated using a Markov Chain Monte Carlo (MCMC) approach, with likelihood function

$$L = \prod_j \lambda_j(T_j) \, e^{-\Sigma_{\tau < T_j} \lambda_j(\tau)}.$$

The model only considers symptomatic infections that occurred in Castiglione d'Adda between July 1 and October 20, 2023. We assumed that the incubation period for dengue in humans, which is the time interval between infection and symptom onset dates, lasts between 3 and 10 days. Symptomatic cases with symptom onset date not later than one week from the first case were excluded from the likelihood function as they might have acquired the infection from unobserved cases that occurred in the early phase of the outbreak. Only cases confirmed by PCR were considered for sensitivity analysis.

The reconstruction of likely transmission chains was carried out by sampling a specific infector $k_j \in I_j$ for each locally infected individual $j$ from a multinomial distribution with probabilities defined as

$$p(k_j, j) = \frac{\lambda_{j,k_j}(T_j)}{\lambda_j(T_j)}.$$

The transmission distance between a case $j$ and its infector $k_j$ is computed after the reconstruction of the transmission chain, by selecting the geographical distance between the residences of the two individuals. The reconstruction of the transmission episodes is repeated 50,000 times to account for model uncertainty.

**Estimation of the reproduction number**

To estimate R(t), we use the same methodology presented in [1,4,5]. We assumed that the daily number of new autochthonous dengue cases (by date of symptom onset) with infection acquired in Castiglione d'Adda L(t) can be approximated by a Poisson distribution according to the equation

$$L(t) \sim \text{Pois}\left( R(t) \sum_{s=0}^{t} \varphi(s)\, C(t-s) \right),$$

where

- $C(t)$, with $t$ from 1 to $T$, is the daily number of new cases (by date of symptom onset) residing

in Castiglione d'Adda;

- $R(t)$ is the net reproduction number at time $t$;
- $\varphi(s)$ is the distribution of the generation time calculated at time $s$, which is assumed to follow a Gamma distribution with mean 18.3 and standard deviation 8.1, as estimated from the data through the approach described in the previous section.

The likelihood $L$ of the observed time series of cases from day 1 to day $T$ conditional on $C(0)$ is thus given by

$$L = \prod_{t=1}^{T} P\left(C(t); R(t) \sum_{s=1}^{t} \varphi(s)\, C(t-s)\right),$$

where $P(k; \lambda)$ is the probability mass function of a Poisson distribution (i.e., the probability of observing $k$ events if these events occur with rate $\lambda$). The posterior distribution of $R_t$ is estimated by using the MCMC Metropolis-Hastings sampling approach. The posterior distribution of $R_0$ is estimated by applying the above-described procedure and by assuming that during the period where the epidemic showed exponential growth $R_t = R_0$ (Figure S1).

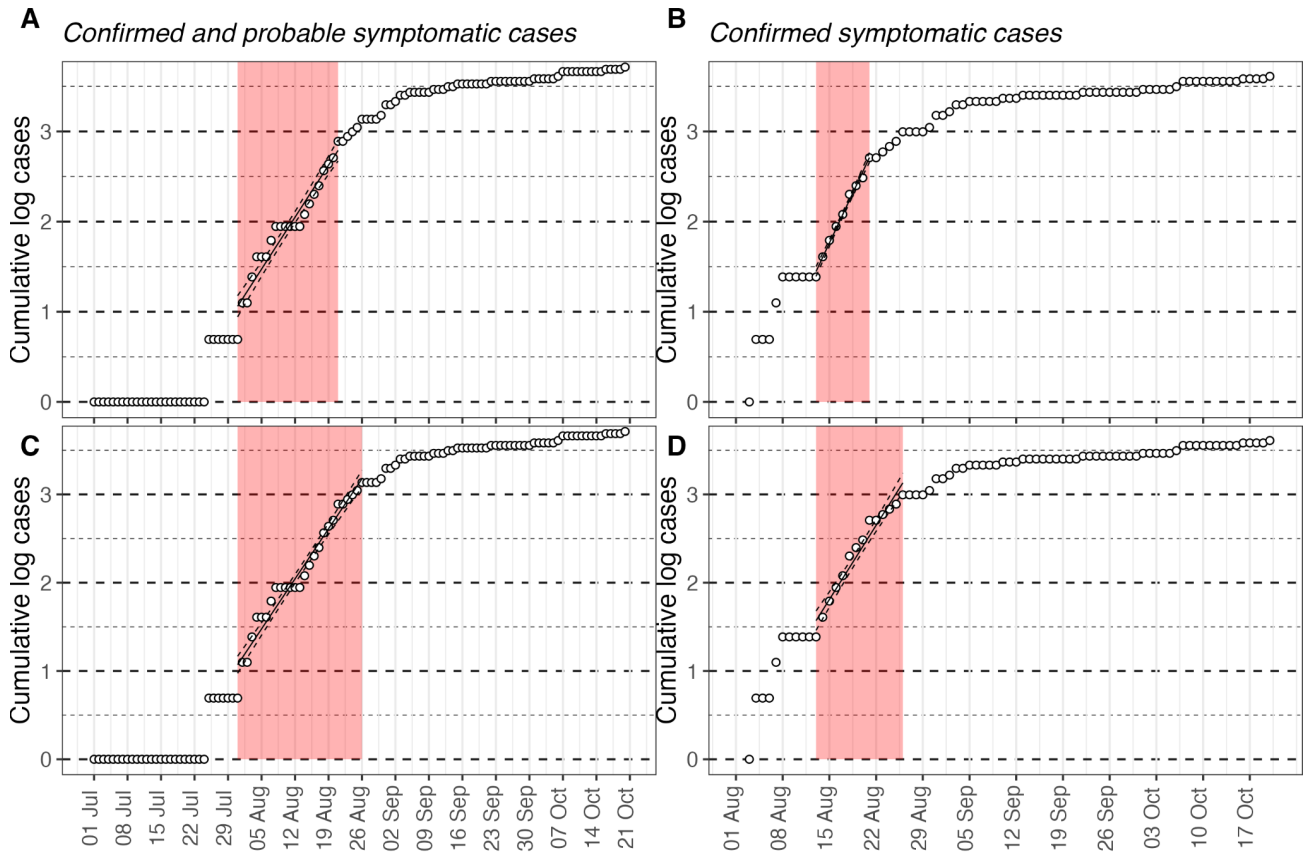

**Figure S1. A** Identified time window (shaded area) of exponential growth in the cumulative number of cases for the estimation of $R_0$. On the x-axis the date, on the y-axis the log-transformed number (plus one) of the cumulative number of daily cases by date of symptom onset. Points represent the log-transformed cumulative number of daily cases infected in Castiglione d'Adda. The solid line represents the regression slope fitted on the log-transformed cumulative number of cases; the dashed line represents the 95% confidence interval. **B** As **A** but excluding cases that were not confirmed by PCR. Panels **C** and **D** show alternative time windows considered for sensitivity analysis.

**Additional results and sensitivity analyses**

The generation time when considering only symptomatic cases confirmed via PCR was estimated to be 17.5 days (95% CI: 12.7-21.6 days). This result is in excellent agreement with what obtained in the baseline analysis (see Figure S2). Considering only cases confirmed via PCR, the average distance between the residences of infector-infectee pairs was estimated to be 279 m (95% CI: 213-348 m). The corresponding cumulative distribution of the transmission distance is shown in Figure S3.

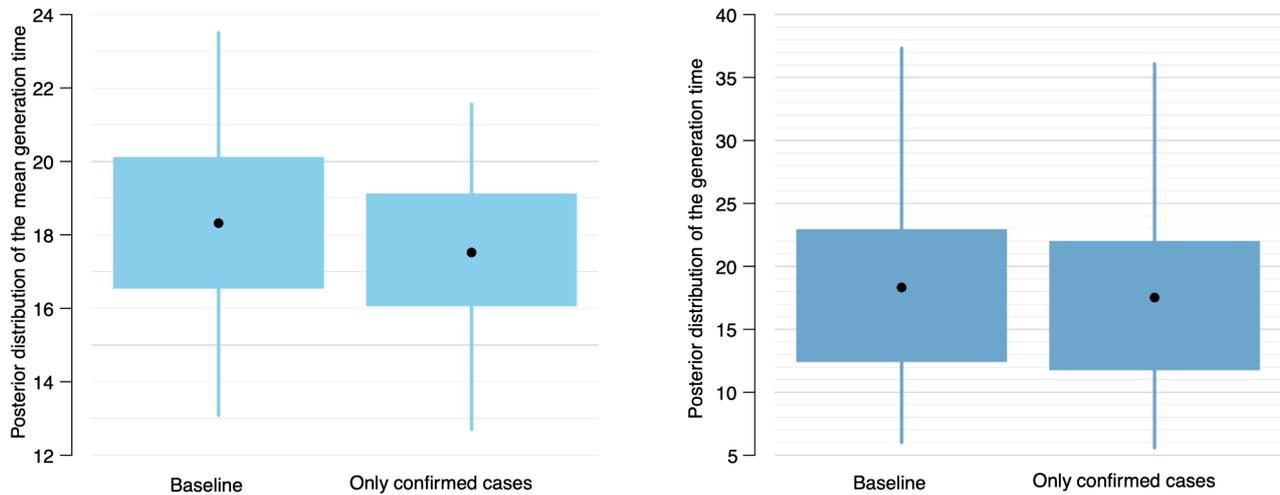

**Figure S2.** Estimates of the generation time distribution, comparing the estimates obtained in our baseline analysis with those obtained when only cases confirmed via PCR were considered. The left panel shows the posterior distribution of the mean generation time, the right panel shows the mean posterior distribution of the generation time. Point: mean value; box: interquartile range; whiskers: 95% CI.

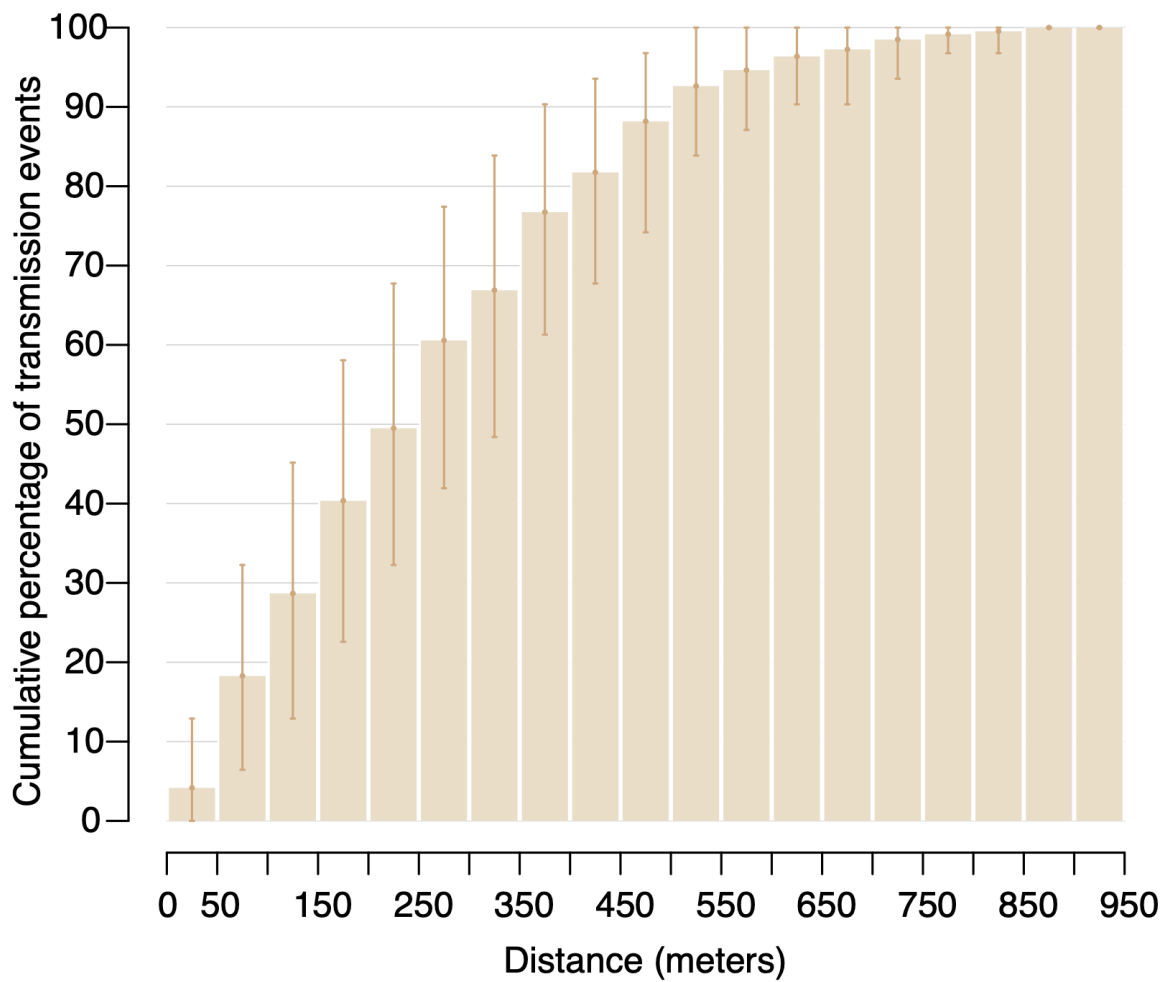

**Figure S3.** Cumulative percentage of the transmission events with respect to the geographical distance between infector-infectee residences when considering only cases confirmed via PCR. Bars: mean value; whiskers: 95% CI.

In the baseline analysis, estimates of the reproduction numbers were obtained by considering all symptomatic cases ascertained in Castiglione d'Adda, and the mean posterior distribution of the generation time (Figure S2). Estimates of the basic reproduction number were obtained by considering the exponential growth of cases observed between July 31 and August 21, 2023 (Figure S1). Alternative estimates of the transmission patterns were explored by considering only cases confirmed via PCR, different time windows for the exponential growth (Figure S1), and both shorter and longer distributions of the dengue generation time. Shorter generation times were explored by considering an average generation time equal to the lower limit of the credible interval of the mean generation time estimated from observed data (namely 13.1 days). Longer generation times were explored by considering an average generation time equal to the upper limit of the credible interval of the mean generation time estimated from observed data (namely 23.5 days).

All the performed analyses provided consistent results on the estimated basic reproduction number (Table S1).

**Table S1.** Estimates of the basic reproduction number $R_0$ as obtained under different model assumptions.

|  |  | *All symptomatic cases* | | *Symptomatic cases confirmed via PCR* | |
|---|---|---|---|---|---|
| *Period of exponential growth* |  | 31 Jul - 21 Aug | 31 Jul - 26 Aug | 13 Aug - 21 Aug | 13 Aug - 26 Aug |
| *Generation time* | Mean | 1.31 (0.76-1.98) | 1.28 (0.80-1.87) | 1.49 (0.78-2.44) | 1.37 (0.80-2.10) |
|  | Shorter | 1.30 (0.75-1.97) | 1.22 (0.77-1.78) | 1.46 (0.76-2.40) | 1.29 (0.75-1.97) |
|  | Longer | 1.31 (0.76-1.99) | 1.30 (0.81-1.89) | 1.49 (0.78-2.45) | 1.40 (0.82-2.15) |

The temporal dynamic of the net reproduction number $R_t$ as estimated by considering all symptomatic cases and alternative distributions of generation time is close to the one presented in the baseline analysis (Figure S4). Similar results were also obtained when considering only cases confirmed via PCR (Figure S5).

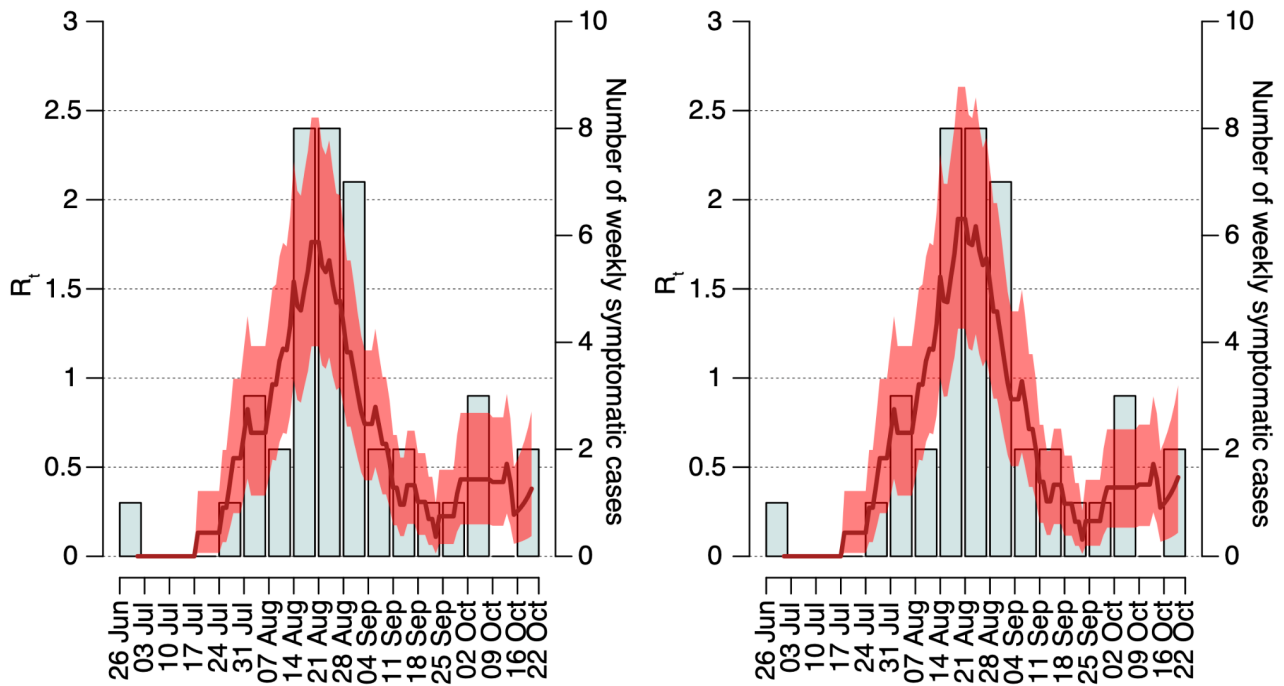

**Figure S4.** Temporal dynamic of the net reproduction number $R_t$ as estimated from all symptomatic cases by considering a shorter (left panel) or longer (right panel) distribution of the generation time. The solid red line represents the estimated mean value of $R_t$; shaded area represents 95% credible interval. Bars represent symptomatic cases aggregated by week of symptom onset (scale on the right y-axis).

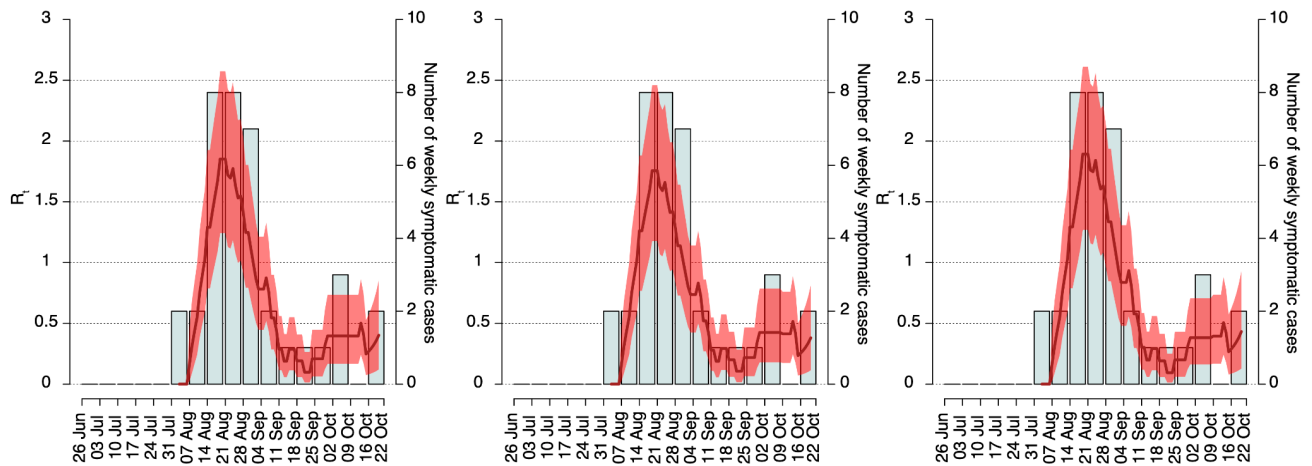

**Figure S5.** Temporal dynamic of the net reproduction number $R_t$ as estimated by considering only cases confirmed via PCR, and using the baseline estimate of the generation time (left panel), or a shorter/longer (central/right panels) generation time. The solid red line represents the estimated mean value of $R_t$; shaded area represents 95% credible interval. Bars represent symptomatic cases confirmed via PCR, aggregated by week of symptom onset (scale on the right y-axis).

**References**

[1] Guzzetta G, Marques-Toledo CA, Rosà R, Teixeira M, Merler S. Quantifying the spatial spread of dengue in a non-endemic Brazilian metropolis via transmission chain reconstruction. Nat Commun. 2018 Jul 19;9(1):2837.

[2] Guzzetta G, Vairo F, Mammone A, Lanini S, Poletti P, Manica M, et al. Spatial modes for transmission of chikungunya virus during a large chikungunya outbreak in Italy: a modeling analysis. BMC Medicine. 2020 Aug 7;18(1):226.

[3] Lau MSY, Dalziel BD, Funk S, McClelland A, Tiffany A, Riley S, et al. Spatial and temporal dynamics of superspreading events in the 2014–2015 West Africa Ebola epidemic. Proceedings of the National Academy of Sciences. 2017 Feb 28;114(9):2337–42.

[4] Cori A, Ferguson NM, Fraser C, Cauchemez S. A New Framework and Software to Estimate Time-Varying Reproduction Numbers During Epidemics. American Journal of Epidemiology. 2013 Nov 1;178(9):1505–12.

[5] Riccardo F, Ajelli M, Andrianou XD, Bella A, Manso MD, Fabiani M, et al. Epidemiological characteristics of COVID-19 cases and estimates of the reproductive numbers 1 month into the epidemic, Italy, 28 January to 31 March 2020. Eurosurveillance. 2020 Dec 10;25(49):2000790.